\documentclass[aps, prl, 
amsmath, showpacs, preprintnumbers, twocolumn, amssymb, amsmath, superscriptaddress]{revtex4-1}
\usepackage{graphicx}
\usepackage{mathptmx, textcomp}
\usepackage[latin1]{inputenc}
\usepackage{tabularx}
\usepackage{units}
\usepackage{color}
\usepackage{bm}

\begin{document}
\title{Revealing the velocity uncertainties of a levitated particle in the quantum ground state}
\author{M.\,Kamba}
\affiliation{Department of Physics, Tokyo Institute of Technology, Ookayama 2-12-1, Meguro-ku, 152-8550 Tokyo}
\author{K.\,Aikawa}
\affiliation{Department of Physics, Tokyo Institute of Technology, Ookayama 2-12-1, Meguro-ku, 152-8550 Tokyo}

\date{\today}


\begin{abstract}
We demonstrate time-of-flight measurements for an ultracold levitated nanoparticle and reveal its velocity for the translational motion brought to the quantum ground state. We discover that the velocity distributions obtained with repeated release-and-recapture measurements are significantly broadened via librational motions of the nanoparticle. Under feedback cooling on all the librational motions, we recover the velocity distributions in reasonable agreement with an expectation from the occupation number, with approximately twice the width of the quantum limit. The strong impact of librational motions on the translational motions is understood as a result of the deviation between the libration center and the center of mass, induced by the asymmetry of the nanoparticle. Our results elucidate the importance of the control over librational motions and establish the basis for exploring quantum mechanical properties of levitated nanoparticles in terms of their velocity. 
\end{abstract}

\maketitle


The ingenious control over the motions of nano- and micro-mechanical oscillators has, over the past decade, opened up a wide variety of opportunities such as quantum transducers~\cite{andrews2014bidirectional,bagci2014optical}, ultrasensitive force and position sensors~\cite{tao2014single,wilson2015measurement}, and nonreciprocal devices\cite{shen2016experimental,bernier2017nonreciprocal,peterson2017demonstration,fang2017generalized}.  
Recent years have witnessed remarkable achievements in manipulating levitated nanomechanical oscillators in the quantum regime~\cite{delic2020cooling,magrini2021real,tebbenjohanns2021quantum,kamba2022optical,piotrowski2023simultaneous,kamba2023nanoscale}, opening exciting possibilities of exploring fundamental physics~\cite{bose2017spin,monteiro2020search,manley2021searching} and macroscopic quantum mechanics~\cite{millen2020quantum,millen2020optomechanics,gonzalez2021levitodynamics}.
 
In previous studies with levitated nanoparticles, precision {\it in situ} measurements of their center-of-mass (COM) position have been a central building block for realizing feedback controls at the quantum level~\cite{delic2020cooling,magrini2021real,tebbenjohanns2021quantum,kamba2022optical}, in analogy with experiments on clamped oscillators. In quantum mechanics, the uncertainty principle imposes a restriction that the position and the velocity cannot be measured simultaneously with infinite precision, dictating the importance of measuring them independently. One of the unique features of levitated nanoparticles is the possibility to let them fly freely by releasing them from the trap and to measure their velocities via time-of-flight (TOF) measurements. Such a scheme has been commonly employed in experiments with ultracold atoms, where the momentum distributions and coherence properties of atomic gases are imaged after TOF expansions and the velocity distributions serve as a standard means for thermometry~\cite{bloch2008many}. For levitated nanoparticles, momentum measurements via TOF have been suggested as one of the promising approaches to realize quantum state tomography of their motional states in the quantum regime~\cite{romero2011optically}. Nevertheless, such measurements for levitated nanoparticles have been reported only at high occupation numbers of $>5000$ in the context of static force sensing~\cite{hebestreit2018sensing}.

\begin{figure}[t]
\includegraphics[width=0.95\columnwidth] {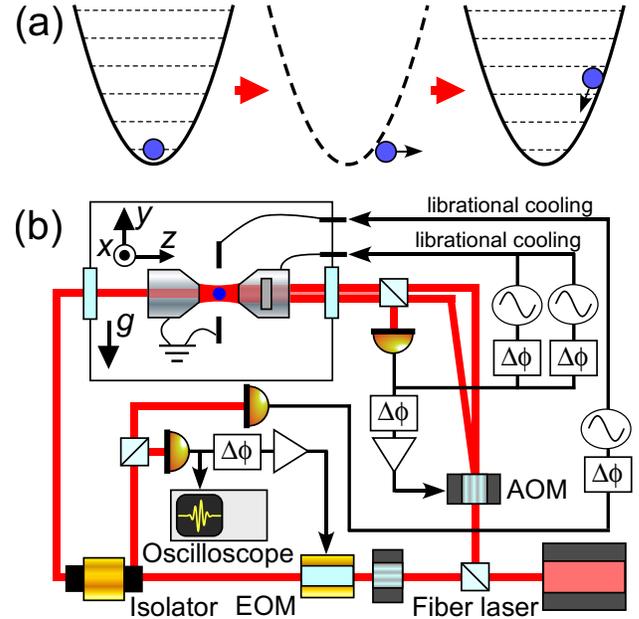}
\caption{ Overview of the experiments. (a) A levitated nanoparticle, whose motional degree of freedom is cooled to the ground state, is released from a harmonic trap and recaptured to the same trap. From the amplitude of the oscillation, the displacement during the TOF is derived. In the presence of librational motions in an optical trap, the nanoparticle can rotate during the TOF. (b) Schematic of the experimental setup. A nearly spherical nanoparticle is trapped in an optical lattice. Three translational motions are feedback-cooled via optical cold damping, while three librational motions are electrically feedback-cooled. To turn off the light for the TOF measurements, an acousto-optic modulator (AOM) is used. Optical cold damping is realized with an electro-optic modulator (EOM) for the $z$ direction and with an AOM for the $x$ and $y$ directions. In the present work, we explore the motions along the $z$ direction with the oscilloscope. }
\label{fig:expset}
\end{figure}

The present work demonstrates TOF measurements with a release-and-recapture protocol for an ultracold levitated nanoparticle [Fig.\ref{fig:expset}(a)], whose motional degree of freedom is cooled to the ground state of a harmonic potential, thereby revealing their velocity distributions with approximately twice the quantum-limited velocity uncertainty. In comparison with the collection of single-shot TOF measurements for separate particles, the release-and-recapture protocol for a specific particle enables us to realize more elaborate exploration of its motional as well as geometrical properties. The presence of librational motions significantly broadens the velocity distributions, showing their strong impact on the dynamics of translational motions during TOF, while we recover the velocity width expected from the independently measured occupation numbers under feedback cooling of librational motions~\cite{pontin2023simultaneous,kamba2023nanoscale}. Based on a simple model of a rigid body, we identify an atomic scale displacement between the COM of the trapped nanoparticle and the center of librational motions as a cause of the observed broadening. Our study reveals a profound relation between translational and librational motions, that has been imperceptible with {\it in situ} position measurements, and shows the necessity of the control over librational motions in velocity measurements. Our work greatly contributes to the interferometry experiments for levitated nanoparticles~\cite{romero-isart2011large,bateman2014near}, where narrow velocity distributions are highly desirable. The presented approach is also valuable as a means to precisely characterize the minute motion of nanoparticles near the ground state, which is nearly obscured by photon shot noise with {\it in situ} position measurements.

In our experiments, we trap a nearly spherical neutral silica nanoparticle with a radius of $R=\unit[174(3)]{nm}$ and a mass of $m=\unit[4.9(3)\times 10^{-17}]{kg}$ in a single site of an optical lattice in the vacuum chamber (Fig.~\ref{fig:expset}). To form the optical lattice, we focus a single-frequency fiber laser with a wavelength of $\unit[1550]{nm}$ and a power of $\unit[176]{mW}$ to a beam waist of about $\unit[1.2]{\mu m}$ and retro-reflect approximately a quarter of the incident power via a partially reflective mirror placed in the chamber~\cite{fnote7}. The retro-reflected beam has a beam waist of about $\unit[1.7]{\mu m}$. It is crucial to prepare a neutral nanoparticle because the motion of a charged nanoparticle during the TOF is strongly influenced by fluctuating electric fields~\cite{hebestreit2018sensing}. During the TOF measurements, the background gas pressure is kept at about $\unit[2\times 10^{-6}]{Pa}$. By detecting the scattered light via photodetectors, we observe the three dimensional motions of the trapped nanoparticle. The three translational motions are cooled via optical feedback cooling~\cite{kamba2022optical,kamba2023nanoscale,vijayan2023scalable}. The motions along the $x$ and $y$ directions have oscillation frequencies of $\{\Omega_x,\Omega_y\}=2\pi \times \{\unit[62,74\}]{kHz}$ and are cooled to occupation numbers of $\{n_x,n_y\}=\{6(1),6(1)\}$. In the following discussions, we focus on the motion along the optical lattice ($z$ direction), which has an oscillation frequency of $\Omega_z/2\pi = \unit[209]{kHz}$. In the preset study, $n_z$ can be varied in the range between 0.8 and 40 by controlling the feedback gain. 

\begin{figure}[t]
\includegraphics[width=0.95\columnwidth] {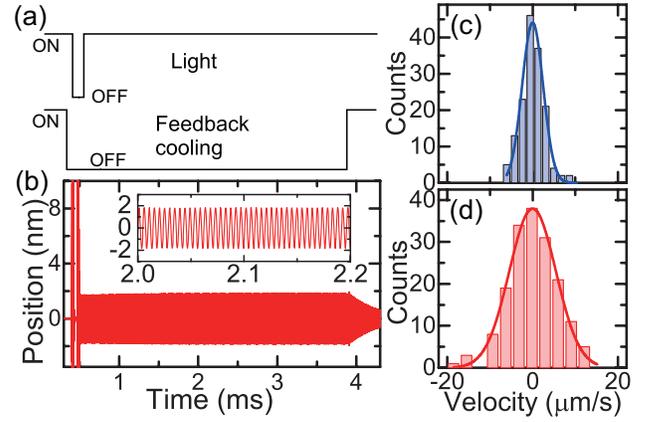}
\caption{ TOF measurements. (a) Time sequence of the TOF measurements. Feedback cooling for translational and librational motions is turned off during the TOF measurements. (b) A typical oscillation signal after the nanoparticle is recaptured. In the inset, an expanded view is shown. The signal is obtained through a high-pass filter such that the motions in $x$ and $y$ directions are excluded from the signal. (c) Velocity distribution for $n_z=0.80$ with LC. The solid line is a Gaussian fit. (d) Velocity distribution for $n_z=0.87$ without LC. The width of the distribution is significantly broadened by the presence of librational motions in the trap. The solid line is a Gaussian fit. }
\label{fig:tof}
\end{figure}

The trapped nanoparticle is slightly deviated from a sphere. In an anisotropic optical trap formed via a linearly polarized light, an aspherical nanoparticle is subject to orientational confinements around three orthogonal axes, resulting in librational motions around these axes~\cite{ruijgrok2011brownian,trojek2012optical,ruijgrok2011brownian,hoang2016torsional,ahn2018optically,bang2020five,van2021sub,kamba2023nanoscale}. These motions have frequencies between $\unit[10]{kHz}$ and $\unit[40]{kHz}$, from which we determine that the nanoparticle is deviated from a sphere by about $\unit[0.5]{\%}$ under an assumption that it is an ellipsoid~\cite{kamba2023nanoscale}. When the nanoparticle is trapped in the optical lattice, the COM motion in the $z$ direction is well-decoupled from librational motions because $\Omega_z/2\pi$ is far from the librational frequencies. Due to the low heating rate of these motions, their amplitudes vary slowly with time scales of more than seconds. The three librational motions can be feedback-cooled by manipulating a naturally existing electric dipole moment in the nanoparticle via time-dependent electric fields synchronized to these motions~\cite{blakemore2022librational,kamba2023nanoscale}. We apply electric fields for librational cooling (LC) via two electrodes and two metal housings for lenses placed in the vacuum chamber (Fig.~\ref{fig:expset}). The temperatures of librational motions are estimated to be lower than \unit[30]{mK}~\cite{kamba2023nanoscale}.

To measure the velocity of the nanoparticle along the $z$ direction, we release the nanoparticle by abruptly turning off the trapping laser for $t_{\rm TOF}=\unit[68]{\mu s}$ and recapture it in the same laser [Fig.~\ref{fig:tof}(a)].  $t_{\rm TOF}$ is chosen to satisfy $\Omega_z t_{\rm TOF} \gg 1$ such that the initial position uncertainty is negligible after the TOF. The nanoparticle is recaptured in the same site of the optical lattice because the displacement during the TOF is more than two orders of magnitude smaller than the lattice spacing. After being recaptured, the nanoparticle oscillates in the optical lattice with nearly constant amplitudes, from which we obtain the position displacements $\Delta z$ during the TOF ~\cite{hebestreit2018sensing} [Fig. ~\ref{fig:tof}(b)]. The position signal is obtained through frequency filters to extract only the COM motions along the $z$ direction~\cite{fnote7}. The velocity of the nanoparticle before the TOF is obtained as $v=\Delta z/t_{\rm TOF}$. Feedback cooling of all the motional degrees of freedom is turned off during this procedure such that it does not affect any motions. 

\begin{figure}[t]
\includegraphics[width=0.95\columnwidth] {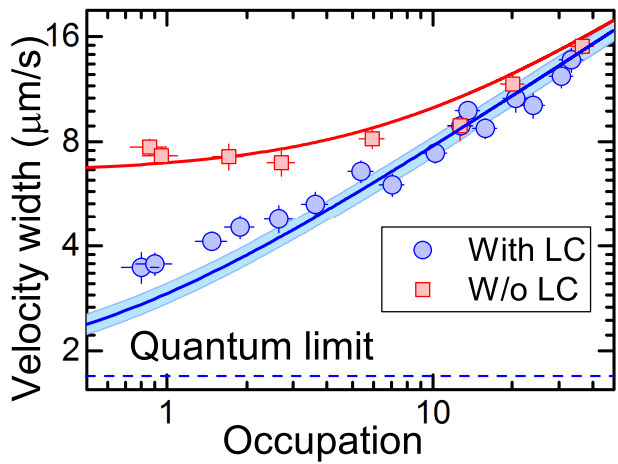}
\caption{Measured velocity width with respect to the occupation number. The vertical error bars reflect both statistical errors in fitting the distributions and systematic errors in calibrating the displacement, while the horizontal error bars indicate systematic errors in temperature measurements. The blue solid line shows calculations with Eq.(\ref{eq:width}) with $\Delta \omega=0$, where the uncertainty in calculations due to the error in $m$ is shown by shaded area. With LC, the observed velocity widths are in reasonable agreement with the calculations. Without LC, the velocity widths are significantly broader than the calculation. The red solid line shows a fit on the results without LC via Eq.(\ref{eq:width}). The dashed line shows the quantum limited velocity uncertainty of $\sqrt{\hbar \Omega_z/m}$. }
\label{fig:width}
\end{figure}

We repeat the same time sequence for about 150 times and derive velocity distributions [Fig.~\ref{fig:tof}(c),(d)]. The number of repetition is confirmed to be sufficient to obtain reliable values for the velocity widths~\cite{fnote7}. The velocity distribution follows the Maxwell-Boltzmann distribution

\begin{align}
f(v) \propto \exp \left(-\dfrac{v^2}{(\Delta v)^2} \right)
\label{eq:fv}
\end{align}
where $\Delta v = \sqrt{\hbar \Omega_z (2n_z+1)/m}$ is the velocity width with $\hbar$ being the reduced Planck constant. Eq.(\ref{eq:fv}) is also valid for a classical harmonic oscillator with the motional temperature of $\hbar \Omega_z (n_z+1/2)/k_{B}$ with $k_{B}$ the Boltzmann constant. The quantum mechanical nature appears as a finite velocity uncertainty of $\sqrt{\hbar \Omega_z/m}$ even at $n_z=0$. The experimentally observed distribution fits well with a Gaussian distribution. From the fit, we extract the width of the distribution which contains the information on the uncertainty in the velocity of the nanoparticle in the optical trap. 

At the lowest $n_z$ close to the ground state, we observe a velocity width of about $\unit[3.5(4)]{\mu m/s}$, which is in reasonable agreement with the value calculated from $n_z$ and is approximately twice the quantum-limited velocity width of $\unit[1.7]{\mu m/s}$ (Fig.~\ref{fig:width}). The slight discrepancy between experiments and calculations at around $n_z \simeq 1$ may suggest the presence of other broadening mechanisms such as fluctuations of retro-reflecting mirrors for forming an optical lattice, which have never been detected with position measurements. When we increase $n_z$, we observe accordingly larger velocity widths, which are in good agreement with the calculated values and confirm the validity of our measurements. 

Surprisingly, we discover that the velocity widths are significantly broadened when LC is not applied. We observe nearly constant velocity widths at $n_z<10$, approximately twice the width obtained at the lowest occupation number with LC. In any cases, the observed profiles are in good agreement with Gaussian. Our observation strongly suggests that librational motions shift the COM position during the TOF randomly. Such a behavior can never occur if the COM of the nanoparticle is placed exactly at the intensity maximum of the optical trap and the center of the librational motions lies at the COM. 

\begin{figure}[t]
\includegraphics[width=0.95\columnwidth] {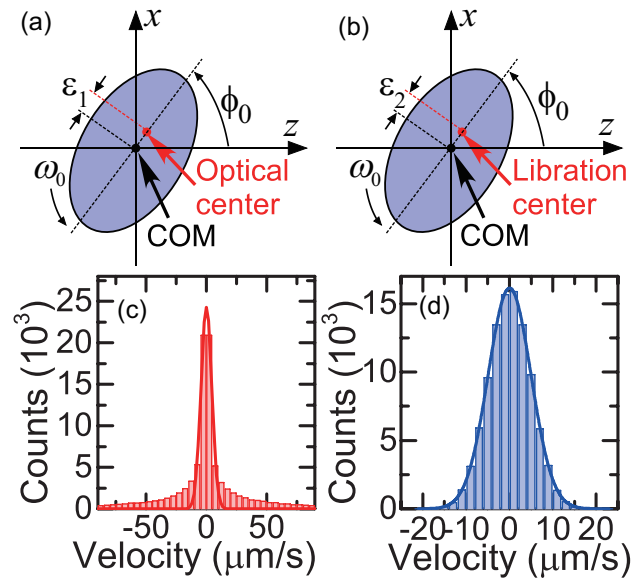}
\caption{Comparison between two models to explain the observed broadening via librational motions. (a) Definition of coordinates for the first model at $t=0$. (b) Definition of coordinates for the second model at $t=0$. (c) Numerically obtained histogram for the first model. The parameter $\epsilon_1$ is set to $\unit[6.7]{nm}$ to reproduce the observed broadening. The solid line is a Gaussian fit. (d) Numerically obtained histogram for the second model. The parameter $\epsilon_2$ is set to $\unit[0.29]{nm}$ to reproduce the observed broadening. The solid line is a Gaussian fit.  }
\label{fig:nums}
\end{figure}

To understand the observed behavior, we consider two possible models of a rigid body in two dimensions, without assuming a specific geometry for the nanoparticle, and compare the results [Fig.~\ref{fig:nums}(a),(b)]. In the first model, we assume that optical feedback cooling of the translational motions locks the point inside the nanoparticle, which we call the optical center, at the intensity maximum of the optical trap, while the COM is displaced by $\epsilon_1$ from the optical center. Because the position measurement is performed optically, what we observe in our experiments is the motion of the optical center. Any librational motion modulates the COM position in the trap even with translational cooling. In this model, the COM has both the translational and angular velocities, both of which induce the displacement of the optical center during the TOF~\cite{fnote7}.

In the second model, we assume that the COM lies at the intensity maximum of the optical trap, while the center of the librational motions, which we call the libration center, is displaced from the COM by $\epsilon_2$. What we optically observe is the motions of the COM. Within the trap, the COM is modulated by librational motions and possesses a finite velocity, which induces the displacement during the TOF. The difference from the first model is the absence of the contribution from the rotation during the TOF~\cite{fnote7}.

We numerically evaluate the distributions obtained with the two models and find a qualitative difference in the profiles [Fig.~\ref{fig:nums}(c),(d)]. The first model exhibits a profile with a long tail at large displacements. In addition, depending on the value of $\phi_0$, the profile is asymmetric~\cite{fnote7}. By contrast, the profile of the second model is symmetric and is Gaussian. From this argument, we conclude that the observed correlation between librational motions and the translational displacement during the TOF is well described by the second model. In the second model, we can derive an expression of the velocity width as

\begin{align}
\label{eq:width}
\Delta v = \sqrt{\hbar \Omega_z (2n_z+1)/m+2\epsilon_2^2 (\Delta \omega)^2}
\end{align}
where $\Delta \omega$ is the uncertainty in the angular velocity before the TOF and $\phi_0=\pi/2$ is assumed for simplicity~\cite{fnote7}. By fitting the observed velocity widths without LC using Eq.(\ref{eq:width}), we find $\epsilon_2 \Delta \omega = \unit[4.4(3)]{\mu m/s}$, from which we derive the displacement $\epsilon_2=\unit[2.0(1)\times10^{-10}]{m}$, comparable to the size of an atom. Here we used a measured uncertainty in the angular velocity of $\Delta \omega/2\pi = \unit[3.5(2)]{kHz}$ due to librational motions around the $x$ and $y$ axis in the absence of LC~\cite{fnote7}. Our model allows us to predict the residual influence of librational motions on the velocity width to be about $\unit[1]{\%}$ of the quantum limit at a librational temperature of $\unit[30]{mK}$.

\begin{figure}[t]
\includegraphics[width=0.95\columnwidth] {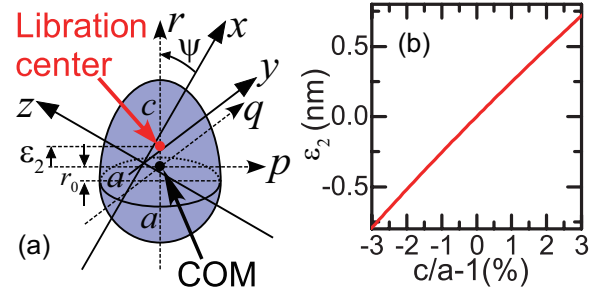}
\caption{ Deviation of the libration center from the COM. (a) Definition of coordinates for an asymmetric nanoparticle to derive the libration center.  We consider an asymmetric nanoparticle obtained by attaching a semisphere (lower half; radius of $a$) and a prolate semi-spheroid (upper half; major semi-axis $c$ and minor semi-axis $a$) at the $r=-r_0$ plane. The COM lies at the origin of the $pqr$ coordinate. The $xyz$ coordinates are defined by the optical trap (lab frame) and are related to the $pqr$ coordinates by $(p,q,r)=\{-z\cos \psi +(x-\epsilon_2) \sin \psi,y,z\sin\psi+\epsilon_2+(x-\epsilon_2)\cos \psi \}$. The nanoparticle is rotated around an axis parallel to the $y$ axis, which passes through the point $(x,y,z)=(\epsilon_2,0,0)$, by $\psi$. Due to the rotation, the COM does not lie at the origin of the $xyz$ coordinate. The coordinates $(p,q,r)$ are used for integrating the optical potential within the nanoparticle.  The libration center lies at $(p,q,r)=(0,0,\epsilon_2)$.  (b) Calculated $\epsilon_2$ as a function of the asymmetry of the nanoparticle $c/a$.  The deviation between the libration center and the COM is calculated for the geometry shown in (a). The radius $a$ is assumed to be $\unit[174]{nm}$. The COM motional frequencies are set to $\Omega_x/2\pi=\unit[62]{kHz}$ and $\Omega_z/2\pi=\unit[209]{kHz}$. These values are required for calculating $\epsilon_2$ via Eq.(9) in the appendix.  }
\label{fig:deviation}
\end{figure}

In the following argument, we consider an asymmetric geometry for the nanoparticle, where an asymmetry indicates a difference in geometry between two halves of the nanoparticle and is not mere differences in radii of an ellipsoid, and elucidate the origin of the deviation of the libration center from the COM. To capture the essence of the problem, we focus on an asymmetric nanoparticle made of a semi-spheroid in the upper side and of a semisphere in the lower side [Figure~\ref{fig:deviation}(a)]. The libration center is determined as a point around which the torque exerted by the optical potential is symmetric. This fact implies that the libration center does not necessarily agree with the COM. In fact, our calculation based on the integration of the optical potential within an arbitrary volume of the nanoparticle under the generalized Rayleigh-Gans approximation~\cite{stickler2016rotranslational,seberson2020stability,rudolph2021theory,fnote7} reveals a displacement of $\unit[1.2\times10^{-10}]{m}$ for the considered geometry with an asymmetry of $\unit[0.5]{\%}$[Fig.~\ref{fig:deviation}(b)]. Given that the actual nanoparticle can be asymmetric in three dimensions, the presented simple model provides a reasonable explanation on the observed deviation of the libration center. Thus, we find that, if the shape of the trapped nanoparticle is known, the TOF measurements in the absence of LC enable us to quantify the asymmetry of the trapped nanoparticle, which is an effect beyond the assumption of a mere ellipsoid.

In conclusion, we realize velocity measurements via TOF with a release-and-recapture protocol for characterizing the motional properties of ultracold levitated nanoparticles brought to the ground state. The demonstrated measurements of velocity widths can also work as an independent thermometry as has been employed in cold atom experiments. Even in the state-of-the-art experiments, the optically observed motion of nanoparticles cooled to the ground state is nearly masked by photon shot noise. TOF measurements magnify their minute motion, thereby enabling us to clearly find the uncertainty of their velocity. One of the important applications of the TOF measurements is quantum state tomography for the motion of nanoparticles~\cite{romero2011optically}. The presented TOF scheme is also useful for observing the quantum interference of rotational motions~\cite{stickler2018probing,stickler2021quantum}. For the purpose of acceleration sensing with levitated nanoparticles~\cite{monteiro2020force,lewandowski2021high}, the ultimate limit originates from the uncertainty in their position, which might be compressed by preparing mechanically squeezed states~\cite{rashid2016experimental,ge2016single,rakhubovsky2019nonclassical,vcernotik2020strong}. To characterize these states, the demonstrated TOF measurements will serve as a crucial tool.  In addition, TOF measurements enables us to measure transient properties of the motions of nanoparticles, thereby enabling us to elucidate the nonequillibrium dynamics of their motion~\cite{militaru2021kovacs,rademacher2022nonequilibrium,wu2022nonequilibrium}. 

We note that, although the first model presented in Fig.~\ref{fig:nums}(a) does not agree with our observations, it is not trivial whether the COM lies exactly at the intensity maximum of the optical trap. Intuitively, the COM should be located at the intensity maximum because each atom in the nanoparticle is expected to equally contribute to both the mass and the optical potential. However, the inhomogeneity of the amorphous glass material of the nanoparticle, giving rise to fluctuations in both the density and the polarizability~\cite{revesz1972pressure,kakiuchida2004refractive}, can cause a deviation of the COM. By enhancing the sensitivity of the measurements with much more TOF repetitions, such a deviation might be detected as non-Gaussian, asymmetric profiles of velocity distributions. The presented scheme opens avenues towards the precision characterization of levitated nanoparticles in terms of their geometry and their material.

\begin{acknowledgments}
We thank M.\,Kozuma, T.\,Mukaiyama, T.\,Sagawa, and K.\,Funo for fruitful discussions. We are grateful to T.\,Tsuda for his experimental assistance. M.\,K. is supported by  the establishment of university fellowships towards the creation of science technology innovation (Grant No. JPMJFS2112). This work is supported by the Murata Science Foundation, the Mitsubishi Foundation, the Challenging Research Award, the 'Planting Seeds for Research' program, Yoshinori Ohsumi Fund for Fundamental Research, and STAR Grant funded by the Tokyo Tech Fund, Research Foundation for Opto-Science and Technology, JSPS KAKENHI (Grants No. JP16K13857, JP16H06016, and JP19H01822), JST PRESTO (Grant No. JPMJPR1661), JST ERATO-FS (Grant No. JPMJER2204) and JST COI-NEXT (Grant No. JPMJPF2015).
\end{acknowledgments}

\section{Supplementary information}
\subsection{Experimental setup}

A single-frequency laser with a power of $\unit[176]{mW}$ is focused with an objective lens (NA$=0.85$) to a beam waist of around $\unit[1.2]{\mu m}$ and is collimated with another objective lens (NA$=0.77$) in the vacuum chamber. Approximately a quarter of the incident power is retro-reflected via a partial reflective mirror placed in the chamber to form a standing-wave optical trap (an optical lattice). The reflected light has a beam waist of around $\unit[1.7]{\mu m}$. We load nanoparticles by blowing up silica powders placed near the trapping region with a pulsed laser at $\unit[532]{nm}$ at pressures of about $\unit[400]{Pa}$. At around $\unit[350]{Pa}$, we apply a positive high voltage to induce a corona discharge and provide a positive charge on the nanoparticle. Then we evacuate the chamber with optical feedback cooling for the translational motions and neutralize the nanoparticle via an ultraviolet light at around $\unit[2 \times 10^{-5}]{Pa}$~\cite{kamba2022optical}.

\begin{figure}[t]
\includegraphics[width=0.95\columnwidth] {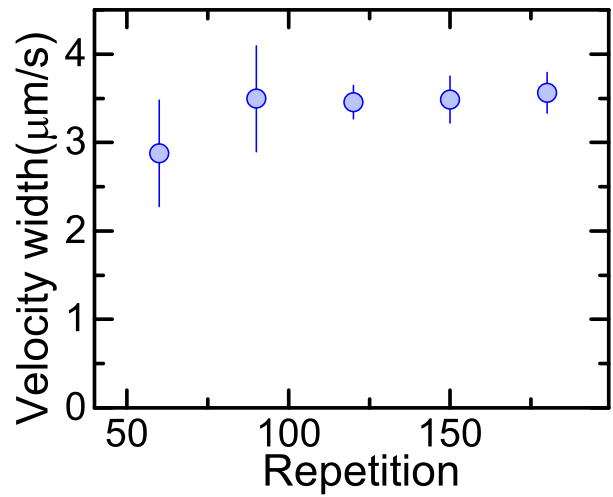}
\caption{ Velocity width with respect to number of data points.  The widths of the velocity distributions are extracted from histograms with various data points. The error bars indicate statistical errors in fitting the distribution.  }
\label{fig:nrep}
\end{figure}

For cooling the translational motion along the $z$ direction, we modulate the optical lattice such that the motion is decelerated (optical cold damping)~\cite{kamba2022optical}. Similarly, we modulate the intensity ratio of the two cooling beams to realize optical cold damping in the $x$ and $y$ directions~\cite{kamba2023nanoscale}.The occupation numbers in each direction are measured by integrating the power spectral density calculated from the position signal~\cite{vovrosh2017parametric,gieseler2012subkelvin,kamba2022optical}.

For cooling librational motions, we apply the feedback signals derived from oscillators phase-locked to the signals from photodetectors (PDs) to electrodes and metal housings for lenses in the vacuum chamber (Fig.~1). The relative phases between the oscillators and the PD are adjusted to achieve maximum cooling efficiencies in each direction. The temperatures of librational motions are lower than $\unit[30]{mK}$ for each degree of freedom~\cite{kamba2023nanoscale}.

\subsection{TOF measurements}

The trapping laser is turned off via an AOM within $\unit[300]{ns}$, which is less than $\unit[10]{\%}$ of the oscillation period along the $z$ direction. Even after the light is turned off, a weak residual light of about $\unit[40]{nW}$ exerts a radiation pressure on the nanoparticle during the TOF and shifts the center of the distribution to about $\Delta z = \unit[2]{nm}$. The center shift does not alter the initial velocity distribution. The residual light mainly originates from the scattering of the trapping laser at optical components. The displacement during the TOF along the $z$ axis is obtained through a high-pass filter at $\unit[150]{kHz}$ and a low-pass filter at $\unit[250]{kHz}$ and is calibrated by assuming that the oscillation amplitude $q_0$ at a pressure of around $\unit[5]{Pa}$ at room temperature $T$ is given by $m \Omega_z q_0^2=2k_{\rm B}T$ with $k_{\rm B}$ being the Boltzmann constant.

\subsection{Two-dimensional models of the influence of librational motions on the COM displacement }
For both models, we assume that, at $t=0$, the COM is at the origin and the angle of the nanoparticle with respect to the $z$ axis is $\phi=\phi_0$. In the first model [Fig.~\ref{fig:nums}(a)], the displacement of the optical center along the $z$ direction during the TOF of $t_{\rm TOF}$ for a single experimental run is given by 

\begin{align}
\Delta z =  v_0t_{\rm TOF} + \epsilon_1 \omega_0t_{\rm TOF}  \sin \phi_0 \notag \\
+ \epsilon_1 \left[ \cos \left(\phi_0+\omega_0 t_{\rm TOF} \right) - \cos \phi_0 \right]
\end{align}
where $v_0$ and $\omega_0$ denote the velocity along the $z$ direction and the angular velocity at the moment of the release from the trap, respectively. In the second model, the displacement of the COM during the TOF for a single experimental run is given by

\begin{align}
\Delta z =  v_0t_{\rm TOF} + \epsilon_2 \omega_0t_{\rm TOF}  \sin \phi_0 
\end{align}

Now we consider the distribution of $\Delta z$ for many experimental runs. For both models, the values of $v_0$ and $\omega_0$  depends on the phases of the translational and librational oscillations, respectively, and therefore vary among experimental runs. The first term, determined by $n_z$, and the second term, determined by librational motions, fluctuate independently. We assume that these two terms have Gaussian distributions, resulting in Eq.(\ref{eq:width}). In the numerical simulation shown in Fig.~\ref{fig:nums}, the value of $\phi_0$ is fixed at $\phi_0=\pi/2$ for simplicity. Because the trapped nanoparticle is expected to possess a preferential orientation determined by its geometry~\cite{ruijgrok2011brownian,trojek2012optical,ruijgrok2011brownian,hoang2016torsional,bang2020five,van2021sub,kamba2023nanoscale}, it is reasonable to assume a fixed value for $\phi_0$.

\begin{figure}[t]
\includegraphics[width=0.95\columnwidth] {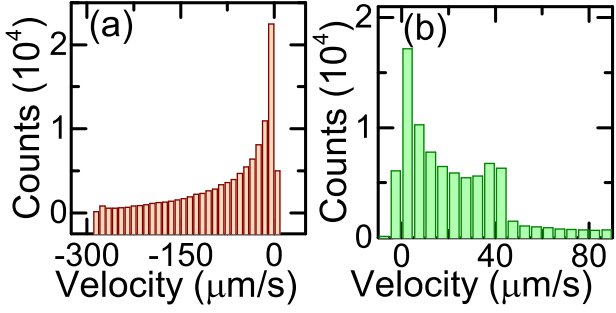}
\caption{ Calculated asymmetric histograms obtained with the first model for variable values of $\phi_0$. (a) $\phi_0=0^\circ$. (b) $\phi_0=135^\circ$.  When the optical center does not agree with the COM, the profiles of velocity distributions strongly depends on the angle and can be asymmetric. Experimentally, we have not observed these profiles. }
\label{fig:twodim}
\end{figure}

\subsection{Measurements of the uncertainty in the angular velocity without LC}
We measure the amplitudes of librational motions by employing the strong nonlinearity of the angular potential~\cite{kamba2023nanoscale}. Due to the nonlinearity, the frequency of the librational motion depends on the amplitude and its variation serves as a thermometry. Because we have no information on which librational motion has a preferential impact on the displacement during the TOF, we take into account both angular velocities around the $x$ and $y$ axes. The measured angular velocities follow the Boltzmann distribution, from which we derive the uncertainty in the angular velocity of $\Delta \omega/2\pi =\unit[3.5(2)]{kHz}$. The librational motion around the $z$ axis does not contribute to the displacement in the $z$ direction.

\subsection{Derivation of the libration center for an asymmetric nanoparticle}
We consider an asymmetric nanoparticle as shown in Fig.~\ref{fig:deviation}(a). The libration center  is displaced from the COM by $\epsilon_2$. To derive an expression for the orientational confinement, we assume that the nanoparticle is rotated by $\psi$ around an axis parallel to the $y$ axis, which passes through the point $(x,y,z)=(\epsilon_2,0,0)$. The oscillation frequencies of the COM around $x,y,z$ axes are $\Omega_x,\Omega_y$, and $\Omega_z$, respectively. Under the generalized Rayleigh-Gans approximation, where we consider the inhomogeneous electric field inside the nanoparticle, the potential energy from the orientational confinement due to the anisotropy of the optical potential is given by~\cite{seberson2020stability,kamba2023nanoscale}

\begin{align}
U=\dfrac{3m}{8\pi a^2(a+c)} \iiint  \Omega_x^2 \left[ r\cos\psi+p\sin \psi +\epsilon_2 -\epsilon_2 \cos\psi  \right]^2\notag \\ +\Omega_y^2q^2
+\Omega_z^2 \left[ p\cos\psi-r\sin\psi+\epsilon_2 \sin\psi \right]^2 dp dq dr
\end{align}
to the lowest order, where $r_0=3(c-a)/8$ and the integration is performed within the arbitrary volume of the nanoparticle. In this argument, we omit the confinement due to the light polarization for simplicity. The confinement due to the light polarization is of the same order as the confinement via the potential ansisotropy~\cite{ruijgrok2011brownian,trojek2012optical,ruijgrok2011brownian,hoang2016torsional,bang2020five,van2021sub,kamba2023nanoscale}.  Including this confinement mechanism in the argument will be an interesting future study. 

We first simplify the potential energy of the nanoparticle at a given libration angle $\psi$ in Eq.(5). By implementing the integrals for $p,q$, we arrive at a simplified representation for an arbitrary range of $r$:

\begin{align}
U(\psi)=&\dfrac{3m}{8(a+c)} \int A(\psi) f^2(r) + \notag \\ &\left[ B(\psi) r^2+C(\psi)r+D(\psi) \right]f(r) dr \\
A(\psi)=&\dfrac{a^2}{4} \left( \Omega_y^2+\Omega_z^2 \cos^2 \psi + \Omega_x^2 \sin^2 \psi \right) \notag \\
B(\psi)=& \Omega_z^2 \sin^2 \psi + \Omega_x^2 \cos^2 \psi \notag \\
C(\psi) =&2\epsilon_2 [\Omega_x^2 \cos\psi- B(\psi)] \notag \\
D(\psi)=&\epsilon_2^2 [\Omega_z^2 \sin^2\psi+\Omega_x^2\left (1-\cos \psi \right )^2] \notag  \\
f(r)=&\left\{ \begin{array}{ll}
1-\dfrac{(r+r_0)^2}{c^2} & (-r_0<r<c-r_0) \notag \\
1-\dfrac{(r+r_0)^2}{a^2} & (-a-r_0<r<-r_0) 
\end{array} \right.
\end{align}

We then calculate the potential energies of the upper half and lower half of the nanoparticle $U_1,U_2$. In the corrdinate of $(p,q,r)$ fixed to the nanoparticle, these volumes correspond to the range of $\epsilon_2<r<c-r_0$ and  $-a-r_0<r<\epsilon_2$. To find the libration center, we require that 

\begin{align}
\dfrac{d^2(U_1-U_2)}{d\psi^2}=0
\end{align}

After cumbersome calculations, we obtain the following representation that determines $\epsilon_2$:

\begin{align}
s \left[ \dfrac{2a^2}{15}(c-a)-\dfrac{a^2}{2}(r_0+\epsilon_2)+\dfrac{a^2}{3c^2}(r_0+\epsilon_2)^3-\dfrac{a^2}{10c^4}(r_0+\epsilon_2)^5 \right. \notag \\ 
\left. -\dfrac{1}{3}(c-r_0)^3 +\dfrac{1}{5c^2}(c-r_0)^5+\dfrac{r_0}{2c^2}(c-r_0)^4+\dfrac{r_0^2}{3c^2}(c-r_0)^3 \right. \notag \\ 
\left. +\dfrac{2}{3}\epsilon_2^3-\dfrac{2}{5c^2}\epsilon_2^5-\dfrac{1}{c^2}r_0\epsilon_2^4- \dfrac{2}{3c^2}r_0^2\epsilon_2^3 -\dfrac{1}{30a^2c^2}r_0^5(a^2-c^2) \right. \notag \\ 
\left. +\dfrac{1}{3}(a+r_0)^3-\dfrac{1}{5a^2}(a+r_0)^5 +\dfrac{r_0}{2a^2}(a+r_0)^4-\dfrac{r_0^2}{3a^2}(a+r_0)^3 \right] \notag \\
+2\epsilon_2 \left(s- \Omega_x^2 \right) \left[ \dfrac{(c-r_0)^2}{2}+\dfrac{(a+r_0)^2}{2}-\epsilon_2^2 \right. \notag \\ 
\left. -\dfrac{1}{c^2} \left \{ \dfrac{(c-r_0)^4}{4} +\dfrac{2}{3}r_0(c-r_0)^3 +\dfrac{r_0^2}{2}(c-r_0)^2 \right \} \right. \notag \\
\left. +\dfrac{2\epsilon_2^2}{c^2} \left( \dfrac{\epsilon_2^2}{4} +\dfrac{2r_0\epsilon_2}{3}+\dfrac{r_0^2}{2} \right) -\dfrac{r_0^4}{12a^2c^2}(a^2-c^2) \right. \notag \\ 
\left. -\dfrac{1}{a^2} \left \{ \dfrac{(a+r_0)^4}{4} -\dfrac{2}{3}r_0(a+r_0)^3 +\dfrac{r_0^2}{2}(a+r_0)^2 \right \}\right] \notag \\
+\epsilon_2^2 \left( 2 \Omega_x^2-s \right) \left[ \dfrac{2}{3}(c-a)-2(r_0+\epsilon_2)+\dfrac{2}{3c^2}(r_0+\epsilon_2)^3 \right] =0
\end{align}
where $s=2(\Omega_x^2-\Omega_z^2)$. By omitting higher order terms of $O(\epsilon_2^3), O(\epsilon_2^2 r_0), O(\epsilon_2 r_0^2), O(r_0^3)$, we arrive at an approximate representation:

\begin{align}
\epsilon_2 = \dfrac{2(\Omega_x^2-\Omega_z^2)(c-a)(19c^2/15-26ac/15+3a^2)}{(\Omega_x^2-2\Omega_z^2)(9c^2+14ac+9a^2)-32a^2(\Omega_x^2-\Omega_z^2)}
\end{align}

\bibliographystyle{apsrev}


\end{document}